%% file: gaz_cms.tex
\documentclass[11pt]{article}
\usepackage{xspace}
\usepackage{graphicx}
\usepackage{amsmath}
\usepackage{amssymb}
\usepackage{color}

\input econfmacros.tex
\def\Title#1{\begin{center} {\Large {\bf #1} } \end{center}}

\begin{document}

\Title{SUSY searches at CMS}

\bigskip\bigskip


\begin{raggedright}  

Alessandro Gaz\index{Gaz, A.}, {\it University of Colorado}\\

\begin{center}\emph{On the behalf of the CMS Collaboration.}\end{center}
\bigskip
\end{raggedright}

{\small
\begin{flushleft}
\emph{To appear in the proceedings of the Interplay between Particle and Astroparticle Physics workshop, 18 -- 22 August, 2014, held at Queen Mary University of London, UK.}
\end{flushleft}
}

\section{Introduction}

The recent discovery of the Higgs boson \cite{Higgs_disc} constitutes a 
magnificent triumph for Particle Physics, but the existence of such elusive
particle also poses difficult challenges for the field. The mass of the Higgs 
boson, not protected by any symmetry, receives quantum corrections from physics 
at higher scales; these corrections, unless some miraculous fine tuning of the 
parameters is in place, are expected to raise the mass of the Higgs Boson to a 
value orders of magnitude higher than the $\sim 126$ GeV that we have observed.

One of the most popular theories that have been proposed to solve this
{\it Hierarchy Problem} is that of Supersymmetry (SUSY), which postulates 
the existence, for each particle of the Standard Model (SM), of a {\it supersymmetric
partner} with spin that differs by 1/2 unity from that of the SM particle.
The existence of these fermion-boson pairs provides a cancellation
mechanism for the corrections of the Higgs mass that greatly reduces the
need for a fine tuning of the parameters. In a {\it natural} scenario, one
in which there is only minimal need for fine tuning, the masses of the
partners of the $t-$ and $b-$quarks, of the gluon, and of the Higgs bosons
are expected to be relatively low, not much above the 1 TeV threshold
(see e.g. \cite{Barbieri}). 

The search for a light $\stop$ has been one of the keynotes of the early SUSY 
searches at the LHC, which focused on $\stop$ pair production with $\stop \ra
t \chiz$ as dominant decay channel. These analyses (see e.g. \cite{early_stop})
already provide tight constraints for the Natural SUSY paradigm, but have little
sensitivity in particular regions of the phase space in which light $\stop$
pairs could still be allowed. One such example is the region in which
$m_{\stop} - m_{\chi} \sim m_{t}$; for this case the stop pair production would not
be experimentally distinguishable from the SM $t \bar{t}$ events. Dedicated 
analyses to cover for these \textit{blind regions} of the classic $\stop$ 
searches have thus been developed and some will be presented here.

In $R-$parity conserving scenarios, the Lightest Supersymmetric Particle (LSP)
is stable and neutral and thus it constitutes a viable candidate for the Dark
Matter in the Universe. In this contributions we focus on these $R-$parity 
conserving scenarios: the stable (and neutral) LSP does not interact with the
detector, thus producing sizable {\it missing transverse energy} ($\MET$). The 
results of the different searches are interpreted both using full realistic
models (pMSSM, cMSSM, mSugra, ... ) and Simplified Models, in which only few
specific production and decay processes for the SUSY particles are considered.
Most of the attention in the last two years has shifted towards the 
interpretation of SUSY results in terms of Simplified Models.

\section{The CMS Detector at the LHC}

The CMS detector is described in detail elsewhere \cite{CMSexp}. Charged tracks
are measured using silicon pixel and strips detectors, the energies of charged and
neutral particles are measured by a lead tungstate electromagnetic calorimeter
and by a brass-scintillator hadronic calorimeter. All these detectors are hosted
inside a superconducting solenoid, providing a magnetic field of 3.8 Tesla. The
flux return yoke is instrumented with gas detectors, that are mainly used for
the detection and momentum measurement of muons. The CMS detector has operated 
reliably during the first years of operations. The results presented here are
based on the full 8 TeV $p-p$ dataset delivered by the CERN LHC in 2012. The
sample corresponds to an integrated luminosity of $\sim 19.5$ fb$^{-1}$.

\begin{figure}[!ht]
\begin{center}
\includegraphics[width=0.55\columnwidth]{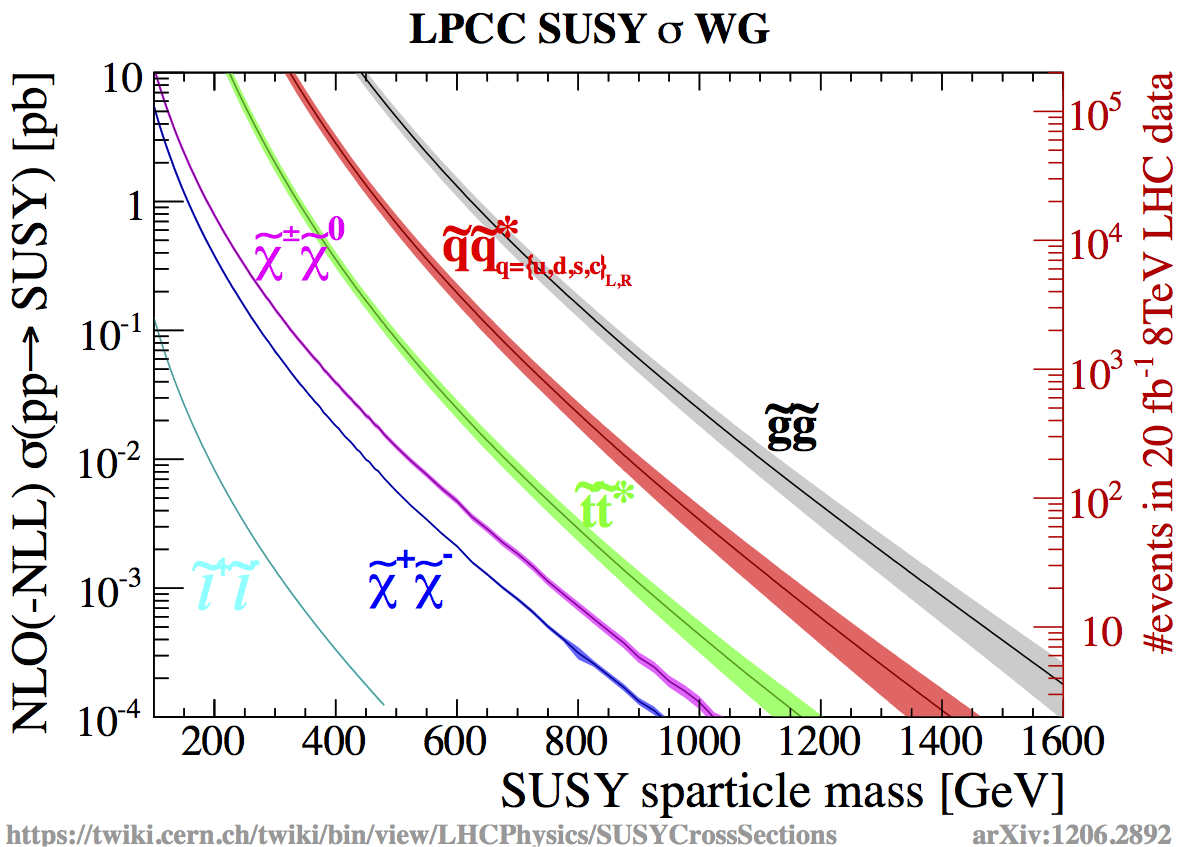}  
\caption{Cross sections for the main SUSY production mechanisms at the
LHC, at a $p-p$ collision energy of 8 TeV.}
\label{fig:Xsecs}
\end{center}
\end{figure}

Fig.~\ref{fig:Xsecs} shows the cross-sections at the LHC for the main production
mechanisms of pairs of supersymmetric particles. The production of strongly
interacting particles can be already probed at masses above 1 TeV. The production
of sleptons, charginos, and electroweakinos is disfavored at a hadron collider,
nevertheless their decays often lead to very clean signatures, so that masses
around 500 GeV could produce a significant signal in the current 
dataset.

\section{Inclusive (hadronic) searches}

\begin{figure}[!ht]
\begin{center}
\begin{tabular}{c c c}
\includegraphics[width=0.31\columnwidth]{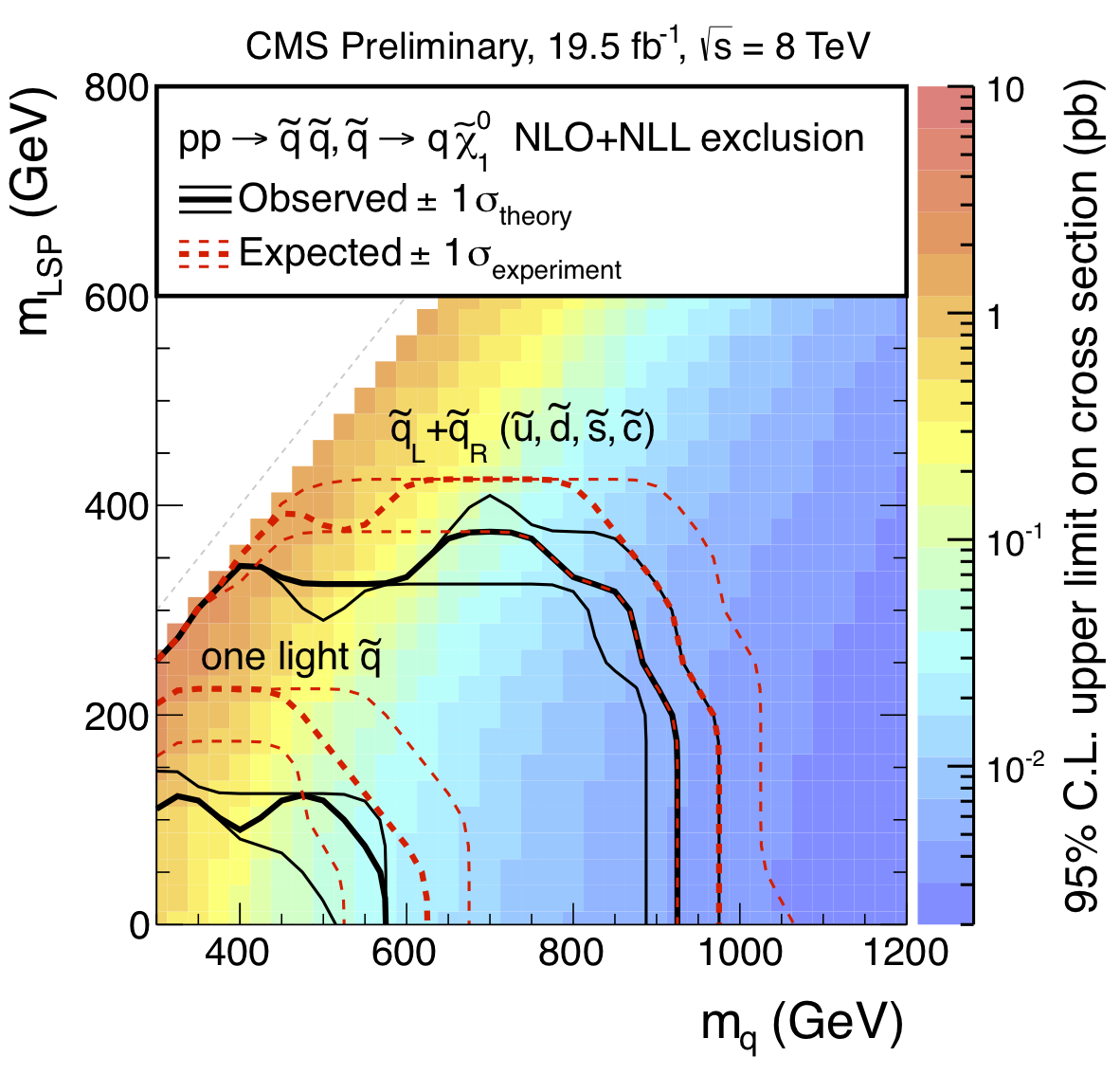} &  
\includegraphics[width=0.31\columnwidth]{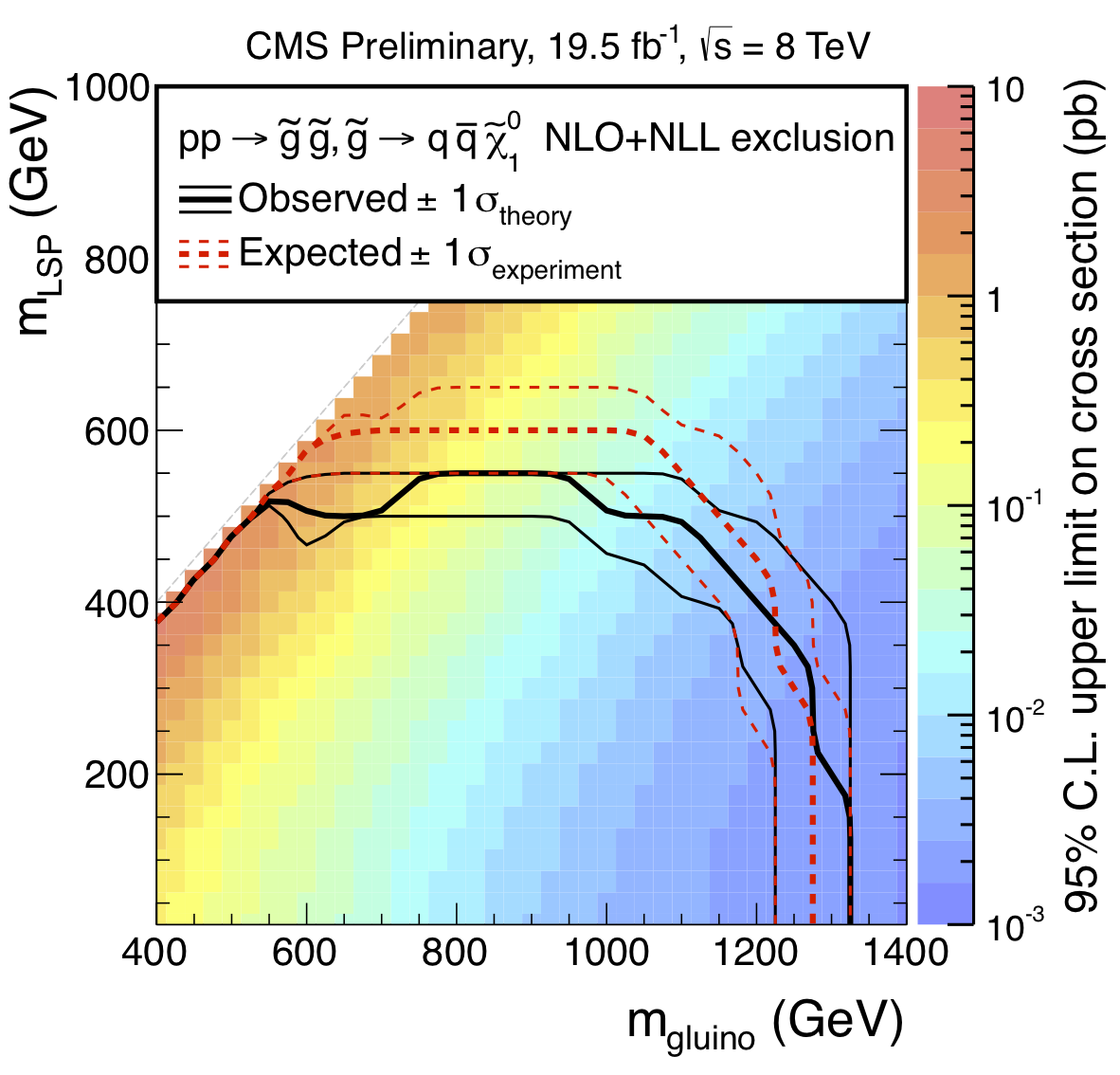} &  
\includegraphics[width=0.31\columnwidth]{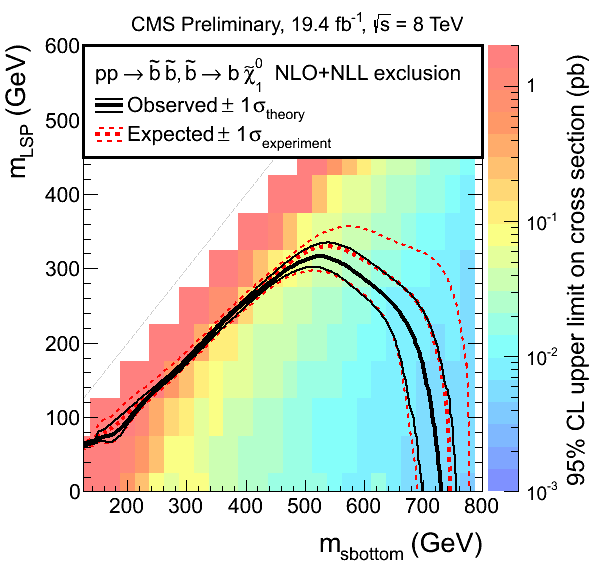} 
\end{tabular}
\caption{Exclusion limits for the CMS $M_{T2}$ analysis \cite{MT2} for direct $\tilde{q}$ pair
production (left plot) and gluino production with each gluino decaying to a light
$q\bar{q}$ pair and the LSP (center plot). Right plot: exclusion limits for the
direct production of $\sbottom$ pairs in \cite{sbottom}.}
\label{fig:MT2}
\end{center}
\end{figure}

Inclusive hadronic analyses target the production of gluinos (that subsequently
decay to a $q\bar{q}$ pair and the LSP, possibly through an intermediate
$q \tilde{q}$ state) and the direct production of squark pairs. Depending on the
specific SUSY model under study, signal events will be characterized by different
jet multiplicity and different $b-$jet multiplicity; the latter is a powerful
tool to discriminate between third-generation SUSY particles (which are expected
to decay dominantly to third generation SM particles) and the first two
generations. CMS searches for these signatures using the $M_{T2}$ variable \cite{MT2}, 
which is defined as:
\begin{equation}
M_{T2} (m_{\tilde{\chi}}) = \min_{\vec{p}_{T}^{\,\tilde{\chi}(1)}+\vec{p}_{T}^{\,\tilde{\chi}(2)} = \vec{p}_T^{\,miss}} \left[ \max \left( M_T^{(1)}, M_T^{(2)} \right) \right] \, ,
\end{equation}
where the minimization satisfies the constraint that the total missing momentum 
is the sum of the missing momentum of the two LSP's present in the event and the
transverse mass $M_T^{(i)}$ is defined as:
\begin{equation}
(M_T^{(i)})^2 = (m^{vis(i)})^2 + m_{\tilde{\chi}}^2 + 2 (E_T^{vis(i)} E_T^{\tilde{\chi}(i)} - \vec{p}_T^{\,vis(i)} \vec{p}_T^{\,\tilde{\chi}(i)}) \, . 
\end{equation}
The $M_{T2}$ variable is designed to be robust against jet energy mismeasurements
that could produce fake $\MET$ in hadronic multijet events. The dataset is subdivided 
into several disjoint signal regions, characterized by different ($b$-)jet multiplicities 
and different values of the scalar sum of the transverse momenta of the jets in the event 
($H_T$). The observed event yields are consistent with the background predictions
that are derived from several data control samples, so exclusion limits are set, using
the signal regions that are expected to be the most sensitive for each model considered.
Some exclusion plots are presented in Fig.~\ref{fig:MT2} (left and center).

Another CMS search \cite{sbottom} targets the direct production of $\sbottom$
pairs, selecting events with exactly two jets with $p_T > 70$ GeV, at least one
of which is identified as a $b$-jet. The discriminating variable is $M_{CT}$,
which is defined as:
\begin{eqnarray}
M_{CT}^2(J_1,J_2) & = & [E_T(J_1) + E_T(J_2)]^2 - [{\bf p_T}(J_1) - {\bf p_T}(J_2)]^2 \\
& = & 2 p_T(J_1) p_T(J_2)(1 + \cos\Delta\phi(J_1,J_2)) \, . 
\end{eqnarray}
The dominant sources of background for this topology are $Z$ and $W$ + jets events,
with $Z \ra \nu\nu$ and $W \ra \ell\nu$; a single $\mu$ data control sample is
utilized to predict their contributions in the signal region.
No significant excess over the predicted background is observed, so upper limits
on the $\sbottom$-pair production are set (see Fig.~\ref{fig:MT2}, right).

\begin{figure}[!ht]
\begin{center}
\begin{tabular}{c c}
\includegraphics[width=0.58\columnwidth]{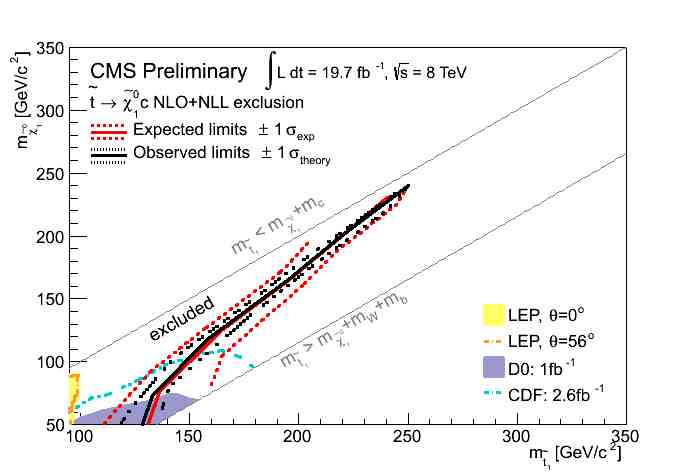} &  
\includegraphics[width=0.40\columnwidth]{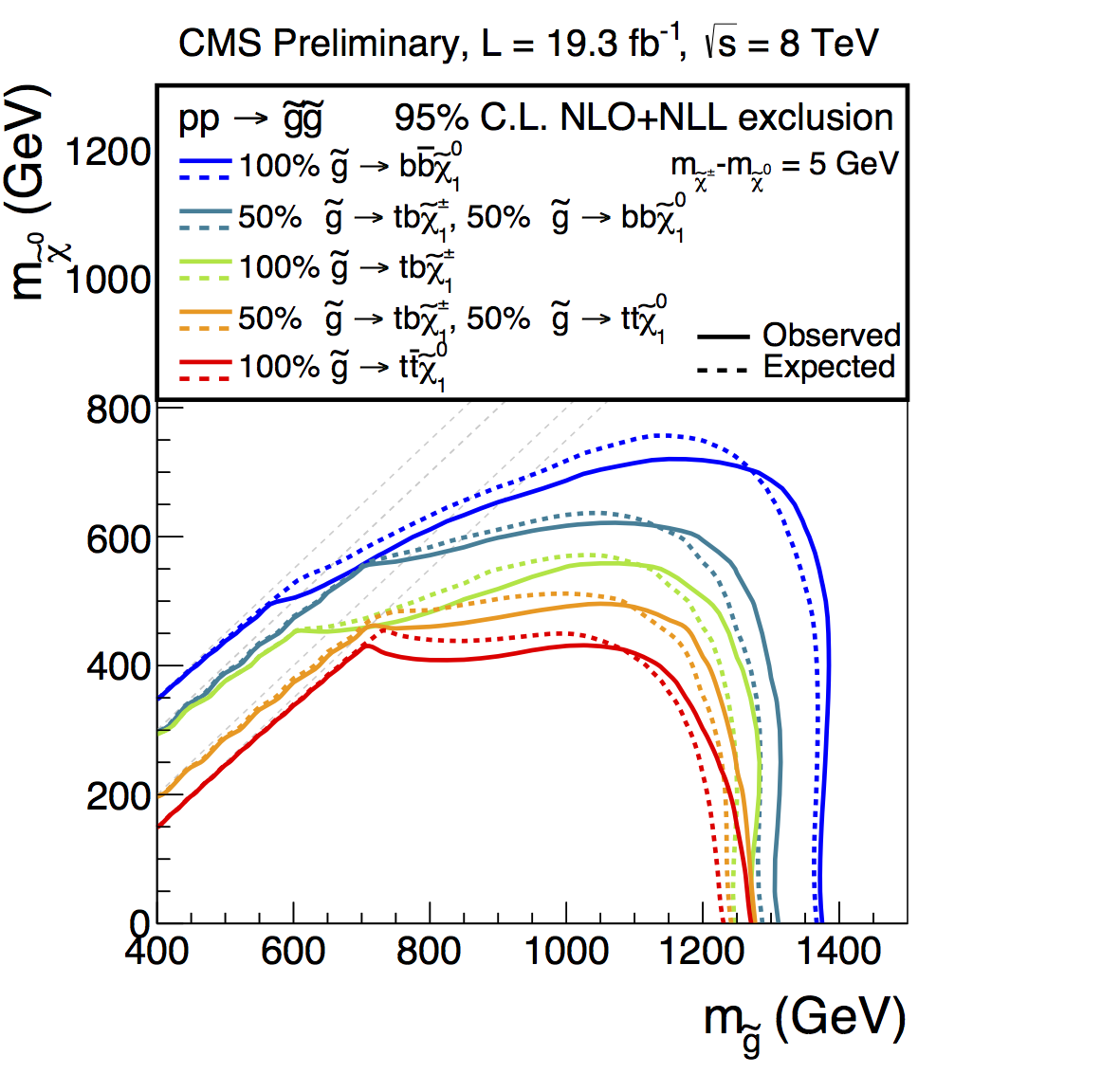} 
\end{tabular}
\caption{Left: exclusion limits for the monojet search \cite{monojet}.
Right: exclusion limits for gluino pairs production, with gluinos decaying
to different mixtures of $b \bar{b} \chiz$, $b \bar{t} \chiz$, and $t \bar{t} \chiz$,
obtained by the razor inclusive analysis \cite{razor_incl}.}
\label{fig:monojet}
\end{center}
\end{figure}

The monojet search \cite{monojet} targets the direct $\stop$-pair production.
As mentioned before, there is a strong belief that the mass of the 
lightest $\stop$ is relatively low in the SUSY spectrum. Initial $\stop$ 
searches focused on the $t\bar{t} + \MET$ signature, excluding direct $\stop$
pair production up to $m_{\stop} \sim 700$ GeV for a light LSP (see e.g. 
\cite{razor_incl} for a combination of a single lepton and a razor inclusive CMS 
analyses). In the case $10 < m_{\stop} - m_{\chi} < 80$ GeV, the dominant decay is expected
to be $\stop \ra c \chiz$. The signature would consist of a soft hadronic jet
associated to a limited amount of $\MET$, that would not offer a handle for an
efficient separation from the SM backgrounds, unless it recoils against a hard
Initial State Radiation (ISR) jet. CMS searches for this signature, selecting
events with a hard ISR jet (and allowing for at most one more soft jet). The
modeling of the ISR is crucial for this analysis and for other analyses targeting
compressed spectra: discrepancies between the data and the simulation are corrected 
for by using high statistics SM control samples ($t\bar{t}$, $Z$+jets, ...). 
Exclusion limits are presented on the left plot of Fig.~\ref{fig:monojet}; the 
sensitivity of the analysis is higher at low values of $m_{\stop} - m_{\chi}$, where 
the events appear to be more monojet-like.

Finally, more exclusion limits on gluino pair production have been obtained by
the razor inclusive analysis \cite{razor_incl}. The branching ratios of the 
gluino decays are varied from the most favorable final state $b \bar{b} \chiz$
to the most challenging $t \bar{t} \chiz$, see Fig.~\ref{fig:monojet} (right).
The difference between the exclusion limits of the two extreme cases (100-150 GeV
for low $\chiz$ masses) gives some feeling about the dependence of the results on
the assumptions on which simplified models are relying.

\section{Kinematic edge in Opposite Sign Dileptons}

Electroweakino decay chains, such as $\tilde{\chi}_0^2 \ra \tilde{\ell} \ell
\ra \ell^+ \ell^- \chiz$, can produce final states containing opposite-charge
same-flavor leptons. The kinematics of this kind of decays is such that the
distribution of the invariant mass of the lepton pair has a {\it triangular}
shape, with a sharp endpoint that is defined by the masses (and the mass
differences) of the particles involved. CMS performed a search for this kind of 
signal in its 2012 dataset, selecting opposite sign electron and muon pairs 
\cite{edge}, and $\MET > 100$ GeV. Events in which both leptons are {\it central} 
($|\eta| < 1.4$) are considered separately from those containing at least a {\it forward} 
lepton ($1.6 < |\eta| < 2.4$). Two strategies are employed for the search: 
\begin{enumerate}
\item{a kinematic fit}, using a signal shape defined by the convolution of
a triangular shape with a Gaussian:
\begin{equation}
\mathcal{P}(m_{\ell\ell}) = \frac{1}{\sqrt{2\pi\sigma_{\ell\ell}}} \int_0^{m_{\ell\ell}^{edge}} y \cdot \exp \left( - \frac{(m_{\ell\ell} - y)^2}{2\sigma^2_{\ell\ell}} \right) dy \, .
\end{equation}
The central and forward regions are fitted simultaneously, with the edge
position $m_{\ell\ell}^{edge}$ being constrained to be the same in the two
regions.
\item{a cut-and-count} analysis is performed in the range $20 < m_{\ell\ell} < 70$ GeV,
independently for central and forward events.
\end{enumerate} 

\begin{figure}[!ht]
\begin{center}
\begin{tabular}{c c}
\includegraphics[width=0.40\columnwidth]{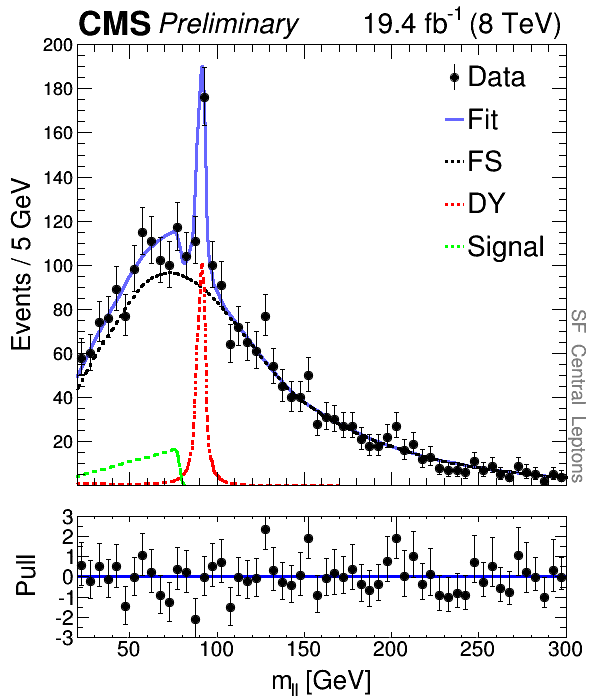} &  
\includegraphics[width=0.39\columnwidth]{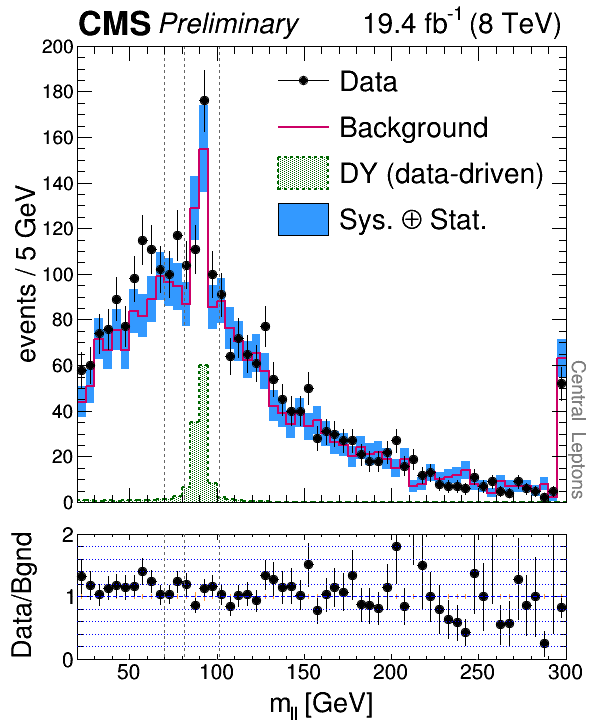} 
\end{tabular}
\caption{Results of the opposite-sign same-flavor dilepton analysis \cite{edge}.
Left: fit of the central events; the black dotted line represents the flavor
symmetric background (dominated by $t\bar{t}$), the red component is the Drell-Yan 
background, and the green dotted line corresponds to the small signal-like excess 
that is seen. Right: results of the cut-and-count analysis for the central events.}
\label{fig:edge}
\end{center}
\end{figure}

The backgrounds can be divided in two main categories: \textit{flavor symmetric}
(that produce same- or opposite flavor dileptonic events with equal
probability) and Drell-Yan (DY), in which only same-flavor pairs are created.
The flavor symmetric background (dominated by $t\bar{t}$ events) is studied
from $e\mu$ events, and its normalization is corrected for the small reconstruction
efficiency differences with respect to the same flavor events. The DY
component is estimated using two (independent) methods: the Jet-Z balance
method (that relies on the imbalance between the $p_T$ of the $Z/\gamma^*$
candidate and that of the hadronic system to separate between SM events
and a potential signal) and the \MET\ template method, in which background
enriched data control samples are used to extrapolate the contribution
at high \MET.

Fig.~\ref{fig:edge} shows the results of the two analysis strategies. Both
approaches find a small signal-like excess, with a significance of 2.4
standard deviations for the kinematical fit and 2.6$\sigma$ (0.3$\sigma$)
for the cut-and-count analysis in the central (forward) region. While the
excess observed is perfectly compatible with a statistical fluctuation of
the backgrounds, this is certainly one of the most interesting hints that
will deserve further scrutiny at the beginning of the LHC Run2.

\section{Searches with photons in the final state}

\begin{figure}[!ht]
\begin{center}
\begin{tabular}{c c}
\includegraphics[width=0.40\columnwidth]{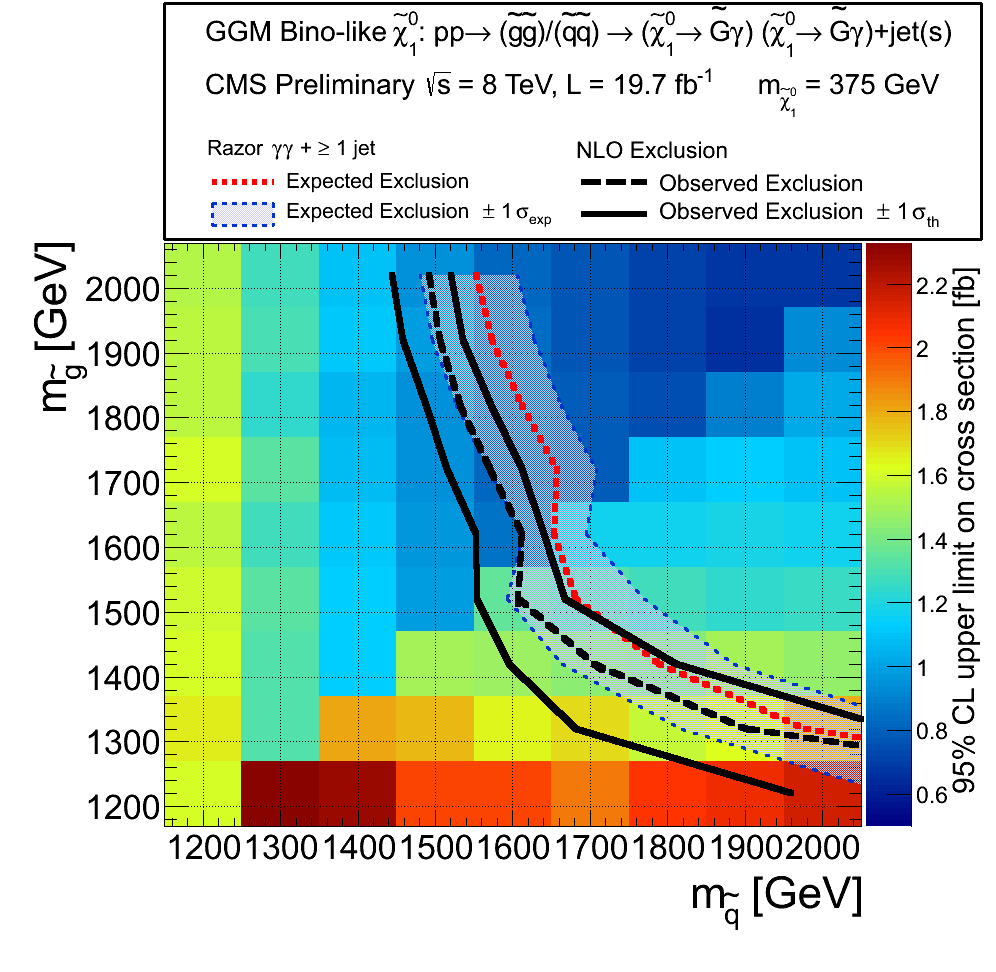} &  
\includegraphics[width=0.40\columnwidth]{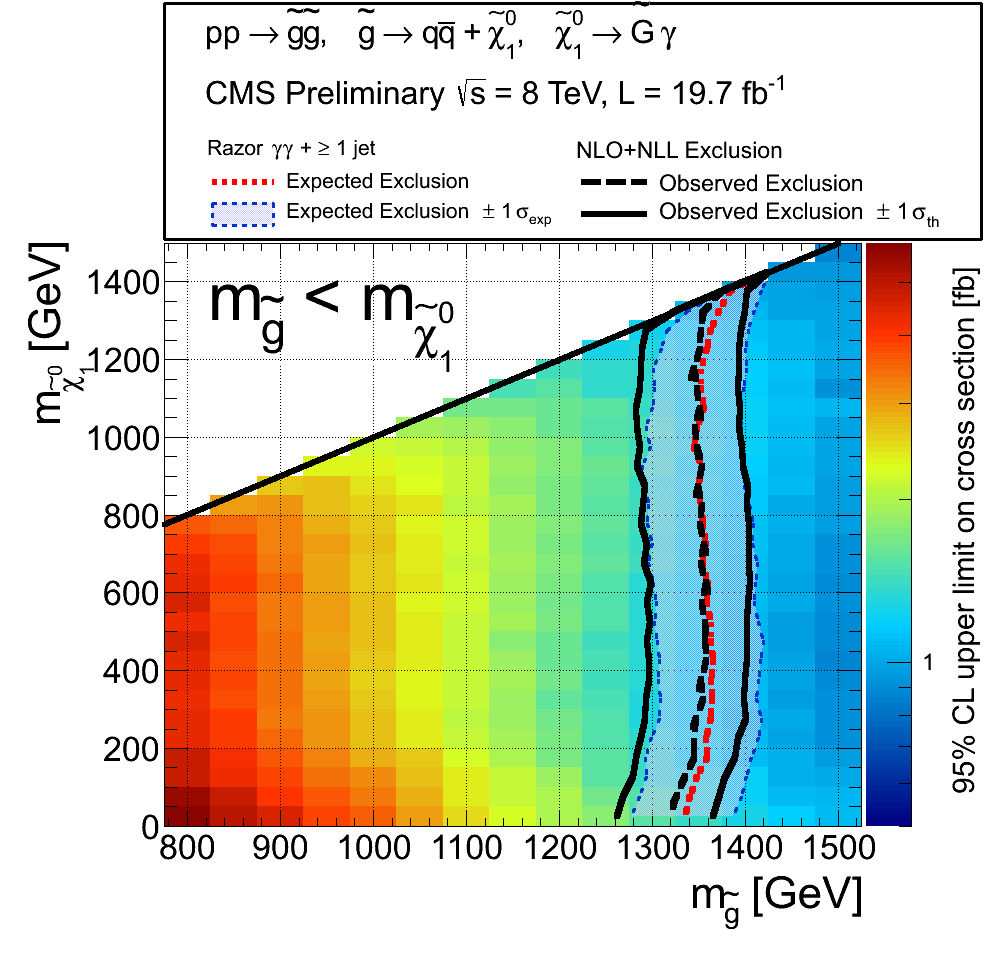} 
\end{tabular}
\caption{Exclusion limits of the razor photon analysis \cite{razor_phot}. Left plot:
exclusion in the $m_{\tilde{g}}$ vs $m_{\tilde{q}}$ plane in a General Gauge Mediated
scenario, right plot: exclusions for the simplified model in which gluino pairs
are produced and each gluino decays to $q\bar{q}\chiz$, with the virtual $\chiz$
decaying to a gravitino $\tilde{G}$ and a photon.}
\label{fig:razor_phot}
\end{center}
\end{figure}

CMS searches for a variety of SUSY models which produce events with at least 
two photons performing a {\it razor} analysis \cite{razor_phot}. The razor 
variables for a dijet event are defined as:
\begin{equation}
\begin{array}{c c}
M_R \equiv \sqrt{(p_{j1}+p_{j2})^2 - (p_z^{j1}+p_z^{j2})^2} \, , & R \equiv M_T^R / M_R \, ,
\end{array}
\end{equation}
where
\begin{equation}
M_T^R \equiv \sqrt{\frac{\MET(p_T^{j1}+p_T^{j2}) - \vec{E}_T^{miss}\cdot(\vec{p}_T^{\,j1}+\vec{p}_T^{\,j2})}{2}} \, .
\end{equation}
For an event containing multiple objects (jets, photons, leptons), the 
objects are combined into two {\it megajets}: the combination which
minimizes the quadratic sum of the invariant masses of the two megajets
is chosen. Events with at least two isolated photons ($p_T >$ 30 and 22 GeV)
and at least one hadronic jet with $p_T >$ 40 GeV are selected. The data 
sample is split into a {\it signal} and a {\it control} region, defined
by the cuts:
\begin{itemize}
\item{signal:} $M_R > 600$ GeV, $R^2 > 0.002$;
\item{control:} $M_R > 600$ GeV, $0.001 < R^2 < 0.002$.
\end{itemize}
For the models under study, the background contamination is negligibly small
in the control region, which is used to fit for the background shape:
\begin{equation}
P(M_R) \propto e^{-k(M_R - M_R^0)^{1/n}} \, ,
\end{equation}
where $k$, $M_R^0$, and $n$ are fit parameters. The results are used to
derive a prediction for the background in the signal region. No significant
excesses are found with respect to the expectations, so limits on a number
of SUSY models are set (see Fig.~\ref{fig:razor_phot}). 

\section{Searches with Higgs bosons in the final state}

\begin{figure}[!ht]
\begin{center}
\begin{tabular}{c c}
\includegraphics[width=0.45\columnwidth]{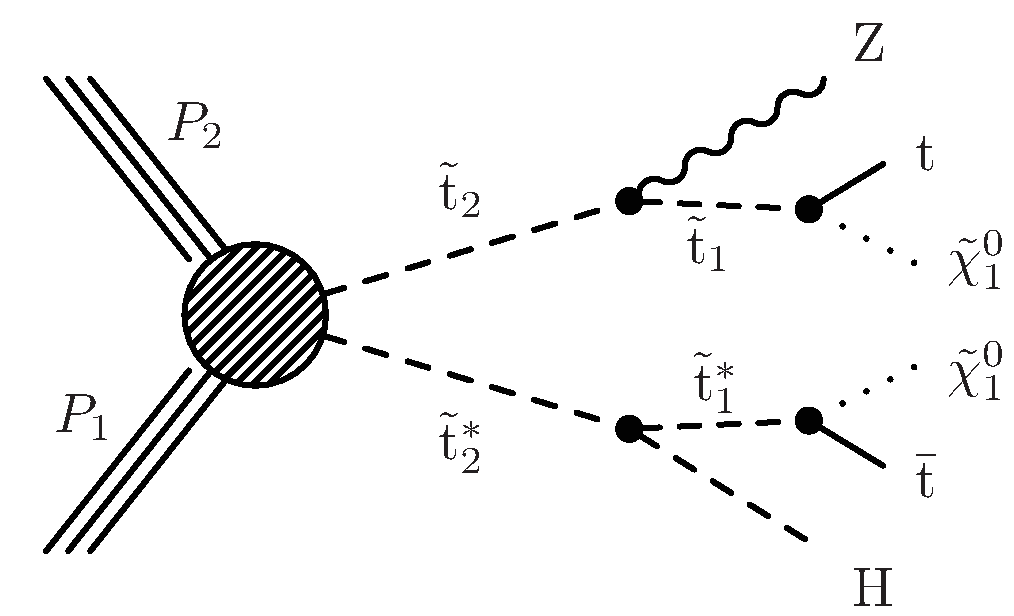} &
\includegraphics[width=0.45\columnwidth]{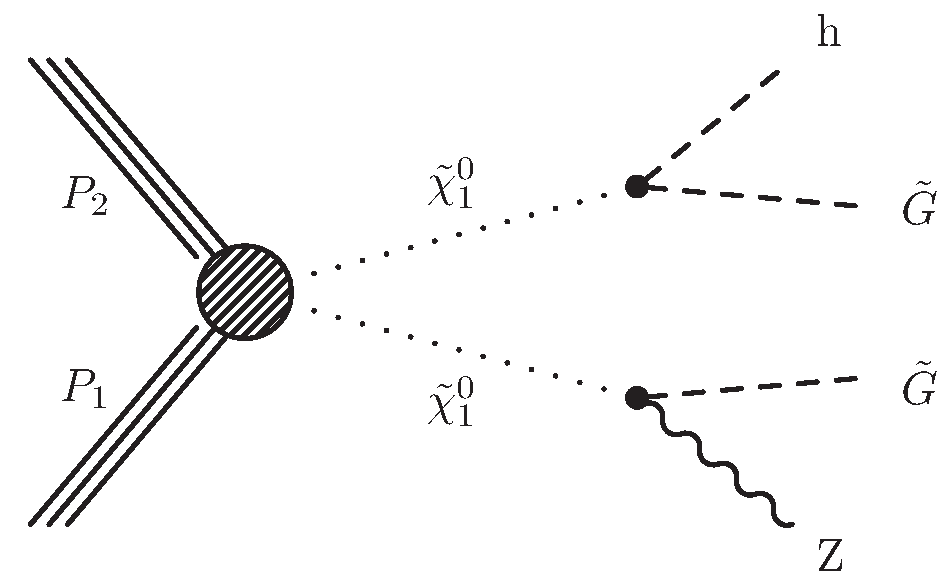} 
\end{tabular}
\caption{Examples of SUSY scenarios producing at least one SM Higgs boson in
the final state. Left: $\stoptwo$ pair production, with $\stoptwo \ra H/Z \, 
\stopone$ investigated in~\cite{higgs_from_stop}; right: higgsino pair
production with $\chi \ra H/Z \, \tilde{G}$ searched for by~\cite{higgs_from_ewkino}.}
\label{fig:higgs_sms}
\end{center}
\end{figure}

In many SUSY models, Higgs bosons are produced in the final state, both from
the decay of strongly interacting sparticles and from electroweakino decays
(see Fig.~\ref{fig:higgs_sms}).

\begin{figure}[!ht]
\begin{center}
\begin{tabular}{c c c}
\includegraphics[width=0.31\columnwidth]{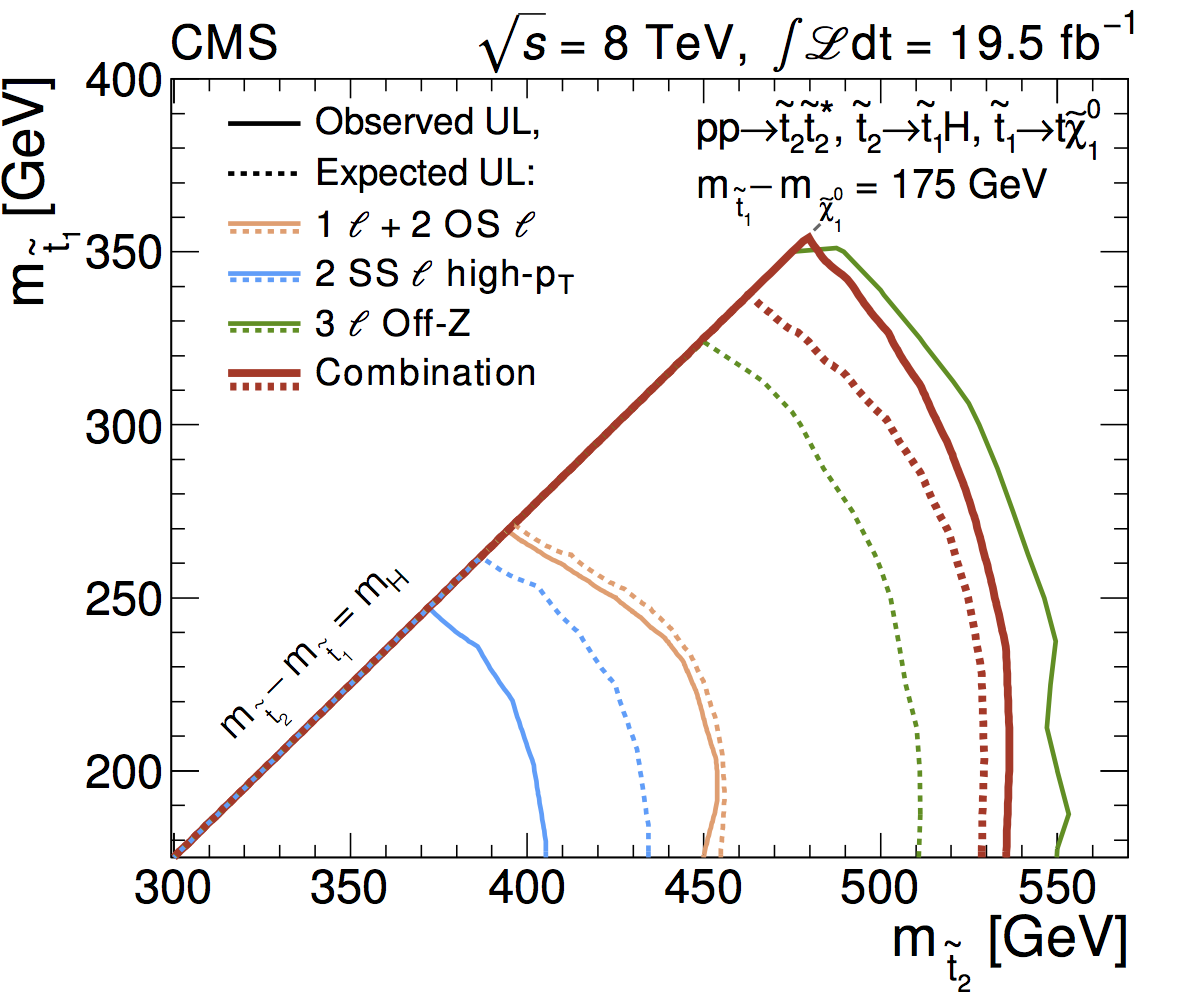} &  
\includegraphics[width=0.31\columnwidth]{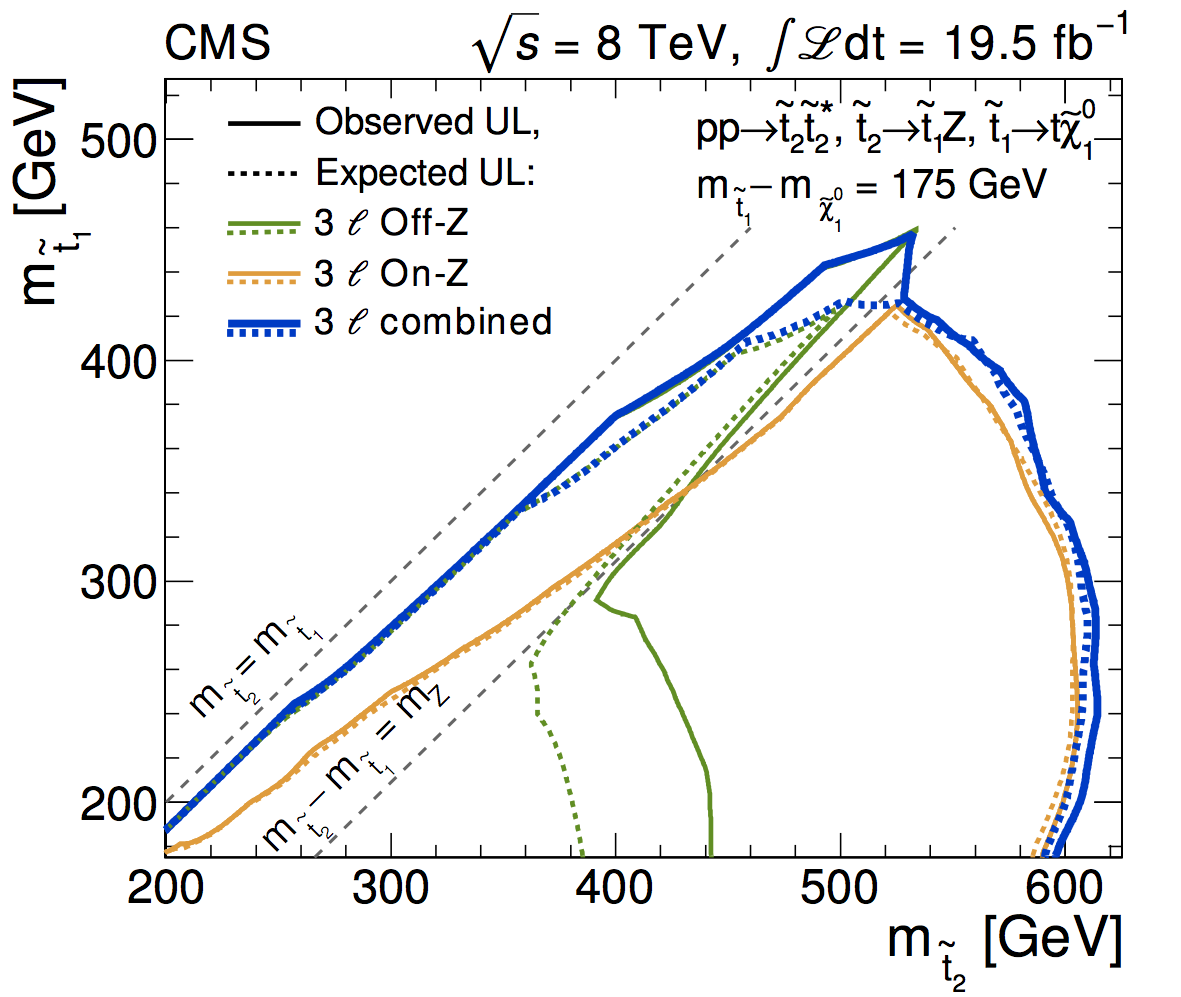} &  
\includegraphics[width=0.31\columnwidth]{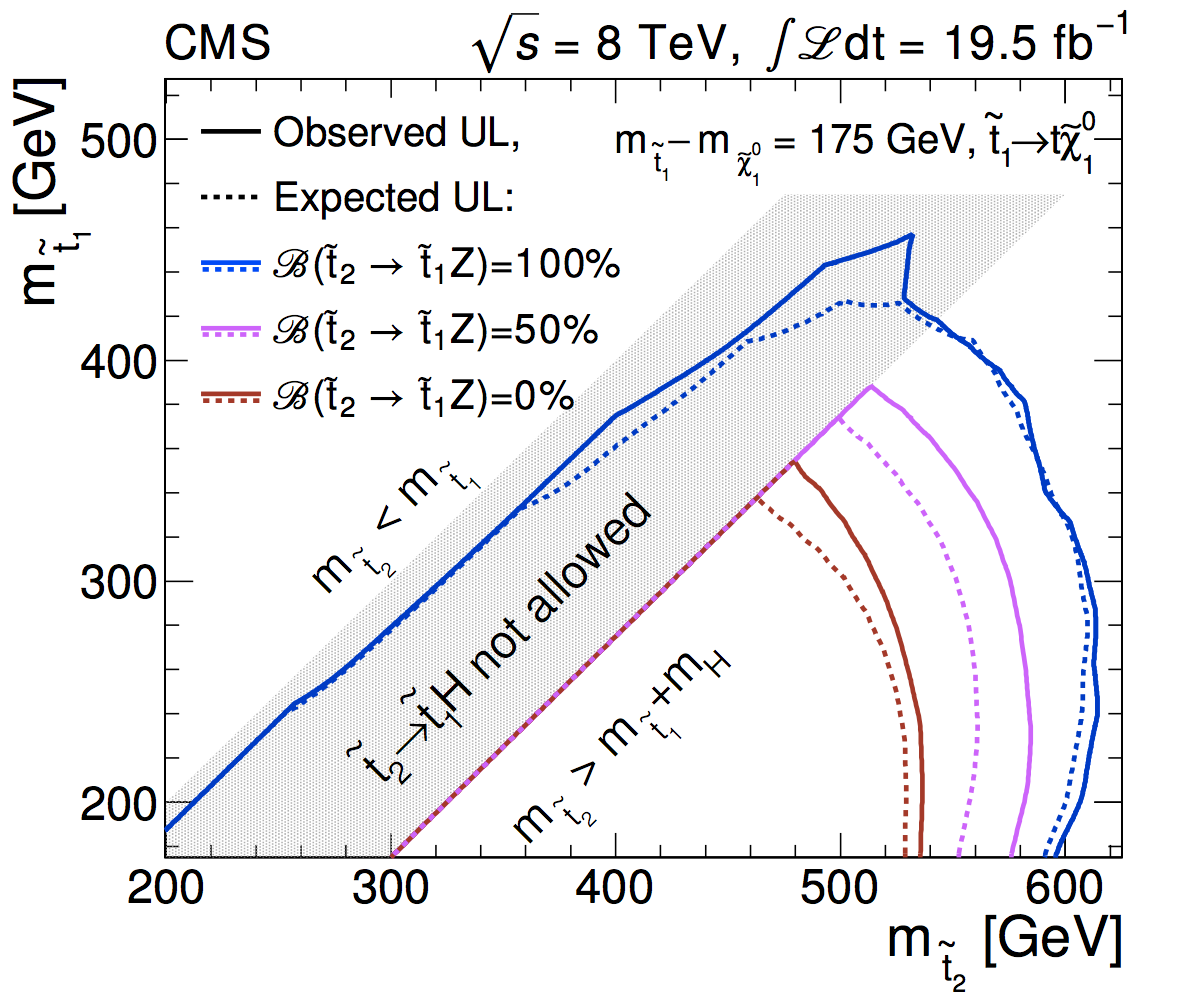} 
\end{tabular}
\caption{Exclusion limits of \cite{higgs_from_stop} for the $\stoptwo \ra H/Z \stopone$ 
simplified model. Left: $\stoptwo \ra H \stopone$ (100\%), center: $\stoptwo \ra Z \stopone$ 
(100\%), right: summary plot with the two extreme cases and the case 
$\BR(\stoptwo \ra H \stopone) = \BR(\stoptwo \ra Z \stopone) =$ 50\%.}
\label{fig:stop2}
\end{center}
\end{figure}

The first kind of scenarios is investigated by CMS to search for $\stoptwo$ production 
\cite{higgs_from_stop}, to target the regions in which the $\stopone$ has a mass 
unfavorable for the classic direct searches. The target model for this analysis
includes $\stoptwo$ pair production, with $\stoptwo \ra H/Z \stopone$, $\stopone
\ra t \chiz$. The branching fraction of $\stoptwo$ to $H$ or $Z$ are varied between
the two extreme scenarios (100\% decays to Higgs bosons or 100\% decays to $Z$'s).
Three search strategies are employed for this analysis, selecting the following 
topologies:
\begin{enumerate}
\item{} events with one lepton or two opposite-sign leptons and at least
three $b$-jets;
\item{} events with two same-sign leptons and at least one $b$-jet;
\item{} events with at least three leptons and at least one $b$-jet.
\end{enumerate}
The dominant SM backgrounds come from $t\bar{t}$ events and diboson ($VV$)
production; these are suppressed by requiring at least one extra lepton
or $b$-jet with respect to the standard topologies, and with additional
requirements on other topological variables like \MET, $M_T(\ell,\MET)$, ...
The observed number of events are consistent with the SM predictions, so
limits on the $\stoptwo$ production are set, see Fig.~\ref{fig:stop2}.
In all cases most of the sensitivity comes from the multilepton channel.

\begin{figure}[!ht]
\begin{center}
\begin{tabular}{c c}
\includegraphics[width=0.48\columnwidth]{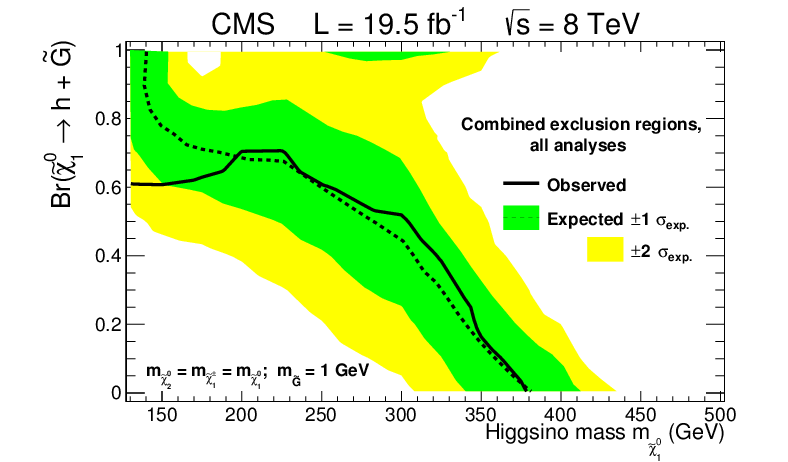} &  
\includegraphics[width=0.48\columnwidth]{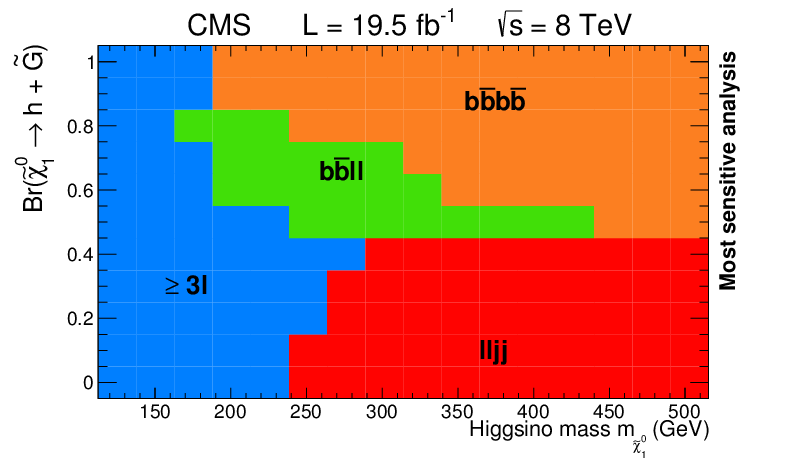}   
\end{tabular}
\caption{Exclusion limits of \cite{higgs_from_ewkino} on higgsino production.
Left plot: limits as a function of the higgsino mass (x axis) and $\BR(\chiz \ra h \tilde{G}) =
1 - \BR(\chiz \ra Z \tilde{G})$ (y axis); the bottom left corner below the black solid line is
excluded by the analysis. The right plot shows the most sensitive channel on the same
plane contributing to the exclusion.}
\label{fig:ewkino}
\end{center}
\end{figure}

The production of electroweakino pairs and their decays to final states containing
Higgs bosons (or $Z$'s and $W$'s) is considered in \cite{higgs_from_ewkino}.
In the target scenario, higgsino ($\chiz$) pairs, or $\tilde{\chi}^0_2 \tilde{\chi}^{\pm}$ 
pairs are produced, with $\chiz \ra h \tilde{G}$, ($\tilde{G}$ is an effectively 
massless gravitino), $\chiz \ra Z \tilde{G}$, $\tilde{\chi}^0_2 \ra h \chiz$, and 
$\tilde{\chi}^{\pm} \ra W^{\pm} \chiz$. Several different channels, each with different 
sensitivity depending on the $\chiz$ mass and on the branching fractions, and associated 
to a significant amount of $\MET$, are considered:
\begin{itemize}
\item{For the $hh$ final state:} $bb \, bb$, $\gamma\gamma \, bb$, $\gamma\gamma \, WW/ZZ/\tau\tau$;
\item{For $hZ$:} $\gamma\gamma \, jj$, $\gamma\gamma \, ee/\mu\mu/\tau\tau$, $bb \, ee/\mu\mu$;
\item{For $ZZ$:} $\geq 3 \ell$, $\ell^+\ell^- \, jj$;
\item{For $hW$:} $\gamma\gamma \, jj$, $\gamma\gamma \, \ell\nu$.
\end{itemize}
No significant excesses over the SM background expectations are found, so 
the channels expected to have significant sensitivity to the model under study
are combined in a likelihood fit. For the higgsino production scenario, with
$\chiz \ra h/Z \tilde{G}$, the limits are evaluated in the $\BR(\chiz \ra h \tilde{G}) =
1 - \BR(\chiz \ra Z \tilde{G})$ vs $m(\chiz)$ plane, see Fig.~\ref{fig:ewkino}. 
For the $\BR(\chiz \ra h \tilde{G}) = 100\%$ case, no exclusion is achieved, even though
the multilepton and $bb \, bb$ analyses get limits within 1$\sigma$ of exclusion at low
and intermediate $\chiz$ masses respectively. In the $\BR(\chiz \ra Z \tilde{G}) = 100\%$
case instead, higgsino pair production is excluded up to $m(\chiz) \simeq 380$ GeV.

\section{Summary}

The CMS Collaboration performed an extensive program of search for the evidence of decays of 
supersymmetric particles, exploiting the full $p-p$ Run1 dataset, collected at center 
of mass energies of 7 and 8 TeV. A variety of search strategies has been employed, ranging
from inclusive searches, sensitive to a wide spectrum of SUSY topologies, to searches
targeting very specific final states. No significant excess has been observed over
the Standard Model backgrounds, so stringent limits are set on potential New Physics
scenarios.

Despite the lack of discoveries in the Run1 data, the motivation for searching for
SUSY signatures at the LHC remains very high as CMS completes the last 8 TeV analyses
and prepares for the beginning of the LHC Run2, scheduled for Spring 2015. The increase to
a center of mass energy of $13-14$ TeV, along with the large integrated luminosity 
that will be collected in the next few years, will allow us to significantly 
extend the sensitivity of SUSY searches and possibly discover or rule out any
Natural SUSY scenario.

What has been presented here is a small selection of results and exclusion limits
produced by the CMS Collaboration; for more information, please see \cite{CMSpublic}.

\bigskip
\section{Acknowledgments}

The author would like to thank the IPA 2014 Organizers for the very interesting
set of topics, the right balance between talks and discussion, and the relaxed
environment in which the Conference took place.

%
%

%
%
%
%
 
\end{document}

%% file: econfmacros.tex
\definecolor{Red}{rgb}{1,0,0}
\definecolor{Green}{rgb}{0,1,0}
\definecolor{Blue}{rgb}{0,0,1}
\definecolor{Black}{rgb}{0,0,0}

\def\beq{\begin{equation}}
\def\eeq#1{\label{#1}\end{equation}}
\def\eeqn{\end{equation}}

\def\beqa{\begin{eqnarray}}
\def\eeqa#1{\label{#1}\end{eqnarray}}
\def\eeqan{\end{eqnarray}}

\let\bar=\overbar

\def\Dslash{\not{\hbox{\kern-4pt $D$}}}
\def\dslash{\not{\hbox{\kern-2pt $\del$}}}

\def\BR{\mbox{\rm BR}}

\def\msb{{\bar{\ssstyle M \kern -1pt S}}}

\def\ra{\rightarrow}
\def\stop{\tilde{t}}
\def\stopone{\tilde{t}_1}
\def\stoptwo{\tilde{t}_2}
\def\sbottom{\tilde{b}}
\def\chiz{\tilde{\chi}^0_1}

\newcommand{\MET}{\ensuremath{E_{\mathrm{T}}^{\text{miss}}}\xspace}

%% file: gaz_cms.bbl
\begin{thebibliography}{99}


\bibitem{Higgs_disc}
G.~Aad {\it et al.}  [ATLAS Collaboration], Phys.\ Lett.\ B {\bf 716}, 1 (2012)
[arXiv:1207.7214 [hep-ex]],
S.~Chatrchyan {\it et al.}  [CMS Collaboration], Phys.\ Lett.\ B {\bf 716}, 30 (2012)
[arXiv:1207.7235 [hep-ex]].

\bibitem{Barbieri}
R.~Barbieri, D.~Pappadopulo, JHEP \textbf{0910}, 61 (2009);
M.~Papucci, J.~T.~Ruderman, and A.~Weiler, JHEP \textbf{1209}, 35 (2012).

\bibitem{early_stop}
S.~Chatrchyan {\it et al.}  [CMS Collaboration], Eur.\ Phys.\ J.\ C {\bf 73}, 2677 (2013)
[arXiv:1308.1586 [hep-ex]].

\bibitem{CMSexp}
S.~Chatrchyan {\it et al.}  [CMS Collaboration],
JINST {\bf 3} S08004 (2008).

\bibitem{MT2}
CMS Collaboration, CMS-PAS-SUS-13-019.

\bibitem{sbottom}
CMS Collaboration, CMS-PAS-SUS-13-018.

\bibitem{monojet}
CMS Collaboration, CMS-PAS-SUS-13-009.

\bibitem{razor_incl}
CMS Collaboration, CMS-PAS-SUS-14-011.

\bibitem{edge}
CMS Collaboration, CMS-PAS-SUS-12-019.

\bibitem{razor_phot}
CMS Collaboration, CMS-PAS-SUS-14-008.

\bibitem{higgs_from_stop}
V.~Khachatryan {\it et al.}  [CMS Collaboration], Phys.\ Lett.\ B {\bf 736}, 371 (2014)
[arXiv:1405.3886 [hep-ex]].

\bibitem{higgs_from_ewkino}
V.~Khachatryan {\it et al.}  [CMS Collaboration], arXiv:1409.3168 [hep-ex], submitted to Phys.\ Rev.\ D.

\bibitem{CMSpublic}
CMS Collaboration,\newline
https://twiki.cern.ch/twiki/bin/view/CMSPublic/PhysicsResultsSUS.


\end{thebibliography}
